\documentclass[12pt]{article}

\usepackage{graphicx}
\usepackage{newtxtext}
\usepackage{newtxmath}
\usepackage{natbib}
\usepackage{hyperref}
\usepackage{tikz}
\usepackage{multirow}
\usepackage{multicol}
\usepackage{pgfplots}
\usepackage{layouts}
\usepackage{comment}
\usepackage{flafter}
\usepackage{float}
\usepackage{authblk}
\hypersetup{
    colorlinks = true,
    urlcolor   = blue,
    citecolor  = black,
}

\newcommand{\RomanNumeralCaps}[1]

\newcommand{\de}[2]{\frac{\partial#1}{\partial#2}}
\renewcommand{\Re}{\operatorname{Re}} 
\renewcommand{\Im}{\operatorname{Im}} 

\title{A theoretical study of the air-sea drag-saturation in very strong winds}

\author[1]{Michael Stiassnie\footnote{Email: miky@technion.ac.il}}
\author[2]{David Andrade\footnote{Email: deandradep@gmail.com}}
\affil[1]{Department of Civil and Environmental Engineering, Technion, Haifa, Israel}
\affil[2]{Centre for Mathematical Sciences, University of Plymouth, Plymouth PL4 8AA, UK}

\begin{document}
\maketitle

\begin{abstract}
The goal of this note is to provide a theoretical explanation for the saturation of the drag coefficient in strong wind conditions. The hydrodynamic model under consideration takes into account the important effects of airborne droplets of water in a thin layer above the water surface that effectively behave as a different fluid between the water and the air. Above this layer the model is coupled with a log-wind profile for the strong winds blowing above the sea. The main underlying mechanism governing the behavior of the drag coefficient is the Kelvin Helmholtz instability for capillary waves on the water surface and the continuity of shear stress along the intermediate interface.
\end{abstract}

\section{Introduction}

The shear-stress between air and sea is an important part of the momentum balance in the development of tropical cyclones.

In moderately wind conditions, with wind speeds less than 20 ms$^{-1}$, it is generally accepted that the drag coefficient increases with wind speed.  However, for hurricane wind speeds (larger than 30 ms$^{-1}$), the experiments of \cite{donelan2004} have convinced the meteorological community that the drag coefficient approaches a limiting value. A property which was basically accepted by the many hundreds of papers citing the above mentioned article.

A few of these papers try to provide some limited theoretical explanation to this phenomenon though it is widely accepted that the influence of water droplets near the water surface is one key factor responsible for the drag reduction. Following \cite{SolovievLukas2010}, droplets are generated by the disruption of the air-sea interface due to wave breaking under strong winds. At wind speeds above 25 ms$^{-1}$, and these droplets effectively form a two-phase layer that covers the sea surface almost completely. A schematic representation is given in figure \ref{fig:1}. Moreover an empirical parametrization of a similar layer, based on the surface wave spectrum, has been proposed by \cite{Iida1992}. 

In the current article, we provide a hydrodynamical explanation based on a linearized formulation for the air-sea system, including effects of airborne water droplets being carried by strong winds above the water surface.  Our model hypothesis is that the droplets effectively create a different fluid between the water and the air characterized by having a different density $\rho_a < \rho < \rho_w$ ($\rho_a$ is the density of the air and $\rho_w$ is the density of the water), that moves with a constant horizontal velocity $U_0$ and has a height $z_0$. This three parameters $\rho$, $U_0$ and $z_0$ are independent and are meant to model different environmental conditions.  When the velocity $U_0$ is above a critical threshold, small capillary waves will appear as a result of the Kelvin-Helmholtz instability. Thus, we shall refer to this two-phase layer as the Kelvin-Helmholtz Layer (K.H. Layer). 

This approach is different from the one proposed by \cite{SolovievLukas2010}, where the influence of the droplets is accounted for by considering a stratified flow with linear velocity and density profiles within a thin strip between the water and the air. A different analytical model, which neglects the influence of the droplets altogether but also explains the drag saturation was proposed by \cite{TroitskayaRybushkina2008}.

Our model contains three different vertical length scales. A small scale for the height of the small capillary waves along the water surface, typically of the order of few millimeters, a small but somewhat larger scale for the thickness of the two-phase layer, typically less than two centimeters, and a much larger scale for the wind profile above the water. As for horizontal length scales we shall be dealing with short capillary waves with typical wavelengths of less than two centimeters, that ride on the much longer waves generated by the storm.

In order to include the wind profile above the water surface we match a log wind profile with the K.H. Layer by requiring continuity of the horizontal velocity and stress between the K.H. Layer and the air. Then, based on the shear-stress, we compute the drag coefficient $C_D$ for different values of $U_{10}$, the wind velocity at a height of $z = 10$ m.

Our model captures the saturation of the drag coefficient for strong winds and offers an explanation as to why it happens; it turns out that $C_D$ strongly depends on the most unstable capillary wave that arises due to the K.H. instability, and reaches breaking conditions.

This paper is organized as follows. Section 2 deals with the basic hydrodynamics of the water and the K.H. Layer. In section 3 we introduce the appropriate wind profile and show how it is coupled with the hydrodynamic model.  Section 4 discusses the drag coefficient and its saturation. Concluding remarks are provided in section 5.

\section{Two Fluid Model}

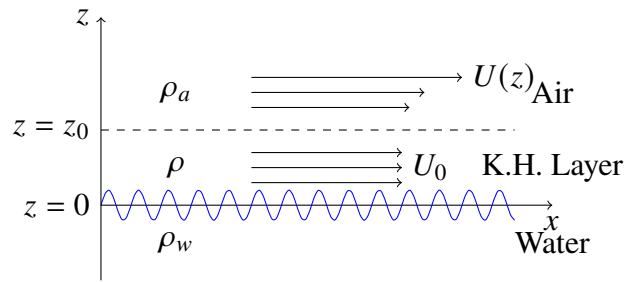
\begin{figure}
\centering
\begin{tikzpicture}
\centering
\draw[->] (0,0) -- (6,0) node[anchor=north] {$x$};
\draw[-,dashed] (0,1) -- (5.5,1);
\draw[-] [domain=0:5.5,samples= 100, smooth, variable=\x, blue] plot ({\x}, {0.2*sin(10*2*pi*\x/4 r )});
\draw[->] (2,0.7) -- (4,0.7);
\draw[->] (2,0.5) -- (4,0.5) node[anchor=west] {$U_0$};
\draw[->] (2,0.3) -- (4,0.3);
\draw[->] (2,1.3) -- (4.1,1.3);
\draw[->] (2,1.5) -- (4.3,1.5) ;
\draw[->] (2,1.7) -- (4.8,1.7) node[anchor=west] {$U(z)$};
\draw	(0,0) node[anchor=east] {$z = 0$};
\draw	(0,1) node[anchor=east] {$z = z_0$};
\draw	(1,-0.5) node[anchor=center] {$\rho_w$};
\draw	(6,-0.5) node[anchor=center] {Water};
\draw	(1,0.5) node[anchor=center ] {$\rho$};
\draw	(6,0.5) node[anchor=center ] {K.H. Layer};
\draw	(1,1.5) node[anchor=center] {$\rho_a$};
\draw	(6,1.5) node[anchor=center] {Air};
\draw[->] (0,-1) -- (0,2.5) node[anchor=east] {$z$};
\end{tikzpicture}
\caption{Schematic representation of the three flow domains.}
\label{model}
\label{fig:1}
\end{figure}

In this section we refer to the two lower domains in figure \ref{fig:1}, namely the water and the K.H. Layer.

Our starting point are the linearized flow equations, (similar to those of  \cite{MikyAgnonJanssen}), for a system of two fluids separated by an interface.

Let $(u_w,v_w)$ be the velocity field of the water, $\rho_w$ its density and $p_w$ the pressure. The equations of motion for the water are 
\begin{align}\label{Euler:water}
&\de{u_w}{x} + \de{v_w}{z} = 0,\quad \text{in $z<0$.}\\
&\de{u_w}{t} = -\frac{1}{\rho_w}\de{p_w}{x},\quad \text{in $z<0$.}\\
&\de{v_w}{t} = -\frac{1}{\rho_w}\de{p_w}{z},\quad \text{in $z<0$.}
\end{align}

Above the water surface, the K.H. Layer is represented by a fluid with constant density $\rho_w < \rho < \rho_a$, moving at constant speed $U_0$ and with height $z_0$. We denote by $(u,v)$ the wavy part of its velocity field  and by $p$ its pressure. The equations of motion are
\begin{align}\label{Euler:layer}
&\de{u}{x} + \de{v}{z} = 0,\quad \text{in $0<z<z_0$.}\\
&\de{u}{t} + U_0\de{u}{x}= -\frac{1}{\rho}\de{p}{x},\quad \text{in $0<z<z_0$.}\\
&\de{v}{t} + U_0\de{v}{x}= -\frac{1}{\rho}\de{p}{z},\quad \text{in $0<z<z_0$.}
\end{align}

Along the mean interface, $z = 0$, we have the kinematic boundary conditions:
\begin{align}
&v_w = \de{\eta}{t},\quad\text{on $z = 0$,}\label{kin1:w}\\
&v = \de{\eta}{t} + U_0\de{\eta}{x},\quad\text{on $z = 0$,}\label{kin1:a}
\end{align}
where $z = \eta(x,t)$ is the interface elevation. Additionally the Bernoulli equation is given by
\begin{align}\label{ber}
p_w - g\rho_w\eta = p - g\rho\eta - \rho_w\tau \frac{\partial^2 \eta}{\partial x^2}.
\end{align}
Here we take $g = 9.81$ms$^{-2}$ for the acceleration due to gravity and $\tau = 7.4\times 10^{-5}$ m$^3$s$^{-2}$ for the ratio between surface tension and the density of the water, which is $\rho_w = 1000$ Kgm$^{-3}$. The density of the air is $\rho_a = 1.2$ Kgm$^{-3}$.

We look for wave-like solutions determined by a wave number $k > 0$ and, in anticipation of instability, a complex frequency $\omega$. The linearized interface is denoted by
\begin{align}\label{eta}
\eta = \Re\left[a_0e^{i(kx - \omega t)}\right],
\end{align}
where $a_0$ is the initial wave amplitude.

We assume that the height of the layer $z_0$ is comparable to the typical wave length of $\eta$, i.e. $\exp(-kz_0)$ is small, thus allowing us to replace the domain of equations \eqref{Euler:layer} from $0 < z < z_0$ by $0 < z$.

In general, one can introduce a velocity potential for each fluid from which one can retrieve the velocity fields and the pressures. The potentials are
\begin{align}
&\phi_w = \frac{a_0}{k}e^{kz}\Im\left[\omega e^{i(kx - \omega t)}\right],\label{phiw}\\
&\phi = -\frac{a_0}{k}e^{-kz}\Im\left[\omega e^{i(kx - \omega t)}\right] + a_0U_0e^{-kz}\Im\left[e^{i(kx - \omega t)}\right].\label{phi}
\end{align}
The pressure of the water is found to be
\begin{align}
p_w = \rho_w\frac{a_0}{k}e^{kz}\Re\left[\omega^2 e^{i(kx - \omega t)}\right],\label{pw}
\end{align} 
and the pressure in the K.H. Layer is 
\begin{align}\label{p}
p = -\rho\frac{a_0}{k}e^{-kz}\Re\left[(\omega - kU_0)^2e^{i(kx - \omega t)}\right].
\end{align} 

Last, substituting equations \eqref{eta}, \eqref{pw} and \eqref{p} in the Bernoulli equation \eqref{ber} yields the following dispersion relation 
\begin{align}\label{omega}
\omega = \frac{\rho_d kU_0}{1 + \rho_d} \pm \left[ \frac{gk(1 - \rho_d)}{1 + \rho_d} + \frac{\tau k^3}{1 + \rho_d} - \frac{\rho_d k^2U_0^2}{(1 + \rho_d)^2}\right]^{1/2} = \frac{\rho_d kU_0}{1 + \rho_d} \pm P^{1/2}(k),
\end{align}
where $\rho_d = \rho/\rho_w$ is the relative density and $P(k)$ is a degree three polynomial, which always has $k = 0$ as a real root.

\subsection{Kelvin Helmholtz Instability Considerations}

Let $U_{cr}$ be the critical speed, given by
\begin{align}
U_{cr} = (1 + \rho_d - \rho_d^2 - \rho_d^3)^{1/4}\left[\frac{4g\tau}{\rho_d^2}\right]^{1/4}.
\end{align}
When $U_0 \leq U_{cr}$, $P(k) \geq 0$ and so $\omega$ is always real. On the other hand  when $U_0 > U_{cr}$,  $P$ has two additional positive roots $k_1$ and $k_2$. For $k_1<k<k_2$, $P(k) < 0$ which leads to a complex frequency $\omega$, implying instabilities with growth rates given by $(-P(k))^{1/2}$.

The critical wave numbers are 
\begin{align}\label{k1,k2}
k_1& = \frac{\rho_d U_0^2 - \sqrt{\rho_d^2U_0^4 - 4g\tau(1 + \rho_d - \rho_d^2 - \rho_d^3)}}{2\tau(1 + \rho_d)}.\\
k_2& = \frac{\rho_d U_0^2 + \sqrt{\rho_d^2U_0^4 - 4g\tau(1 + \rho_d - \rho_d^2 - \rho_d^3)}}{2\tau(1 + \rho_d)}.
\end{align}

The imaginary part of the frequency, i.e. the growth rate, is given by
\begin{align}\label{beta}
\beta = \Im[\omega] = \frac{1}{1 + \rho_d}\left(\rho_d U_0^2k^2 - (1 - \rho_d^2)gk - (1 + \rho_d)\tau k^3\right)^{1/2},
\end{align}
whereas its real part is
\begin{align}\label{alpha}
\alpha = \Re[\omega] = \frac{\rho_d k U_0}{1 + \rho_d}.
\end{align}

Note that in case of instability, all unstable waves with $k_1 < k < k_2$ propagate with the same phase velocity
\begin{align}
c = \frac{\alpha}{k} = \frac{\rho_d U_0}{1 + \rho_d},
\end{align}
which does not depend on $k$, while its amplitude grows exponentially according to
\begin{align}\label{eqn:a}
a = a_0e^{\beta t},
\end{align}
where $a_0$ is the initial amplitude of the wave, as given in \eqref{eta}.

The wave with maximum growth rate, for a given $U_0 > U_{cr}$,  is
\begin{align}\label{km}
k_m = \frac{1}{1 + \rho_d}\left(\frac{\rho_d U_0^2}{3\tau} + \left[\left(\frac{\rho_dU_0^2}{3\tau}\right)^2 - (1 + \rho_d - \rho_d^2 - \rho_d^3)\frac{g}{3\tau}\right]^{1/2}\right).
\end{align}

\subsection{Wave Induced Stress}

Following \cite{phillipsbook}, equation 3.2.11, page 28, the mean momentum per unit area is given by
\begin{align}\label{M:1}
M = -\rho_w\overline{\left(\phi_w(x,0,t)\frac{\partial \eta}{\partial x}(x,t)\right)},
\end{align}
where the bar denotes averaging over one wavelength. This equation is correct to second order. 

By substituting equations \eqref{eta},\eqref{phiw} into equation \eqref{M:1} and averaging, it follows that
\begin{align}\label{M:2}
M = \rho_w\frac{1}{2}\alpha a^2.
\end{align}

Let $\sigma$ be the air-sea shear-stress. This is obtained directly from 
\begin{align}\label{sigma}
\sigma = \frac{d M}{dt} = \rho_w\alpha\beta a^2.
\end{align}

Note that in case of stability, (i.e. $\beta = 0$),  $M$ is constant and there is no shear stress induced by the wave on the water surface.

According to \eqref{eqn:a} any unstable wave can grow indefinitely with time. In reality, an unstable wave will grow exponentially until it reaches its maximum amplitude and breaks. Once the wave breaks it releases water particles that are carried by the wind and join the K.H. Layer. These droplets eventually fall down and coalesce into the water and their momentum is conveyed in the form of stress onto the water surface.

The maximum amplitude that a propagating wave can reach is obtained from 
\begin{align}\label{ak}
a = \frac{\chi(k)}{k},
\end{align}
where $\chi$ denotes the maximum wave steepness. This value is estimated as
\begin{align}\label{chi}
\chi(k)  = 2.292 - \frac{3.630\times 10^2}{k} - \frac{5.017\times 10^5}{k^2} + \frac{2.565\times 10^8}{k^3} - \frac{3.545\times 10^{10}}{k^4}.
\end{align}
This equation was obtained through a polynomial fit of the results of \cite{Hogan1980b} (p. 434). There, the maximum wave steepness is denoted by $h_{max}$ and it is given as a function of his dimensionless parameter $\kappa$.  In equation \eqref{chi}, $\chi \equiv h_{max}$ and $k$ denotes the wave number in m$^{-1}$.

Equation \eqref{chi} recovers the results of Hogan for wave numbers $k =$ 325.65, 364.09, 814.14, 1151.37 m$^{-1}$,  since these were the values used for the polynomial interpolation.  Moreover, as $k$ tends to infinity, $\chi$ recovers the asymptotic value for extremely short waves, given also in \cite{Hogan1980b}.  In this article, all the computations involve waves with $k > 325$ m$^{-1}$ (wave lengths smaller than 1.9 cm), therefore this interpolation is enough for our purposes.

\section{Coupling the K.H. Layer to the wind profile}

So far, our model describes the motion of the water surface under the influence of, mainly, the K.H. Layer above it. In this section we couple a logarithmic wind profile with the K.H. Layer to include the effects of the wind blowing above it.

The criteria to match the horizontal wind profile with the dynamics of the K.H. Layer is that of continuity in velocity and stress. The logarithmic profile that we chose for the wind is 
\begin{align}\label{U}
U(z) = \frac{u_*}{\kappa}\ln\left(\frac{z}{z_0}\right) + U_0, \quad \text{for $z > z_0$.}
\end{align}
Note that this logarithmic wind profile is only defined above $z_0$.  In our coupling, the thickness of the K.H. Layer is taken equal to the wavelength of the most unstable wave $k_m$. Hence 
\begin{align}\label{z0}
z_0 = 2\pi/k_m.
\end{align}

In equation \eqref{U} $\kappa = 0.41$ is the von K\'{a}rm\'{a}n constant and $u_*$ is the frictional velocity which is related to the shear-stress by
\begin{align}\label{u*}
u_* = \left[\frac{\sigma}{\rho_a}\right]^{1/2}.
\end{align}

Once the wind profile is determined, the drag coefficient based on $U_{10}$, the wind velocity at $z = 10$ m, is given by
\begin{align}\label{cd}
C_D = \frac{ u_*^2}{U_{10}^2}.
\end{align}

\section{Saturation of the Drag Coefficient}

\begin{figure}
\centering
\includegraphics[width = \textwidth]{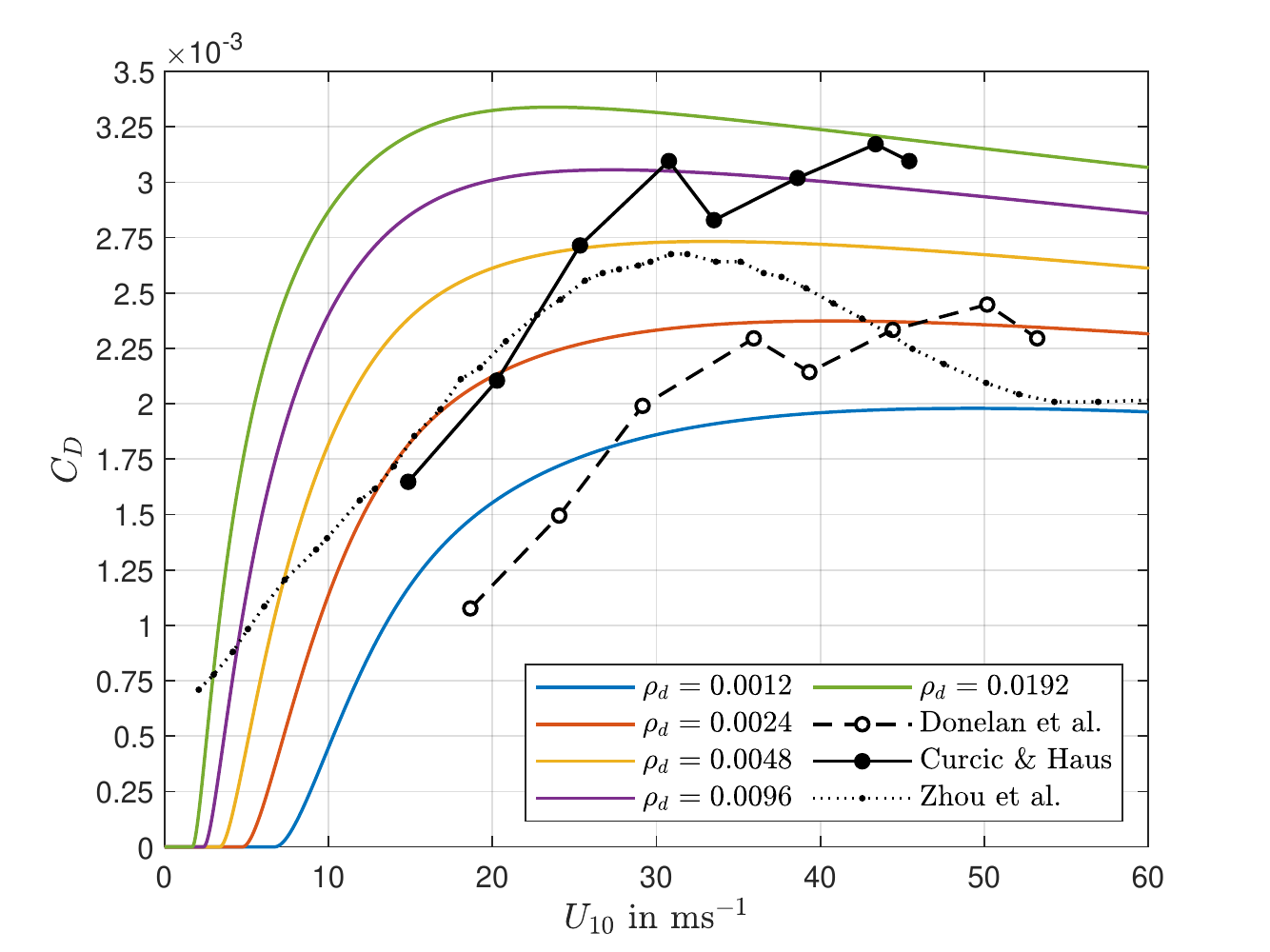}
\caption{Drag coefficient as a function of $U_{10}$, for different values of $\rho_d$. The black broken line shows the original presentation of the measurements taken from  \cite{donelan2004}. The black solid line is the recently corrected presentation of the same measurements as reported in figure 3 of \cite{CurcicHaus}. The doted line shows $C_D$ as implemented in the Geophysical Fluid Dynamics Laboratory hurricane model, reported by \cite{Zhouetal2022}.}
\label{fig:cd}
\end{figure}

Equations \eqref{U} and \eqref{cd} are used as parametric equations of the functional relation between the drag coefficient $C_D$ and the wind velocity at 10 m height $U_{10}$, for a chosen $\rho_d$.

In figure \ref{fig:cd} we plot the drag coefficient as a function of $U_{10}$, for different values of the parameter $\rho_d$. We use $\rho_d = $0.0012, 0.0024, 0.0048, 0.0096 and 0.0192. Note that $\rho_d = 0.0012$ assumes that there are no droplets in the K.H. Layer.

According to our model the typical behavior of $C_D$ can be described as a bounded function of $U_{10}$. It grows monotonically once the velocity inside the K.H. Layer exceeds the critical speed $U_{cr}$, then it  reaches a maximum followed by a slow decrease as the wind speed $U_{10}$ increases. \emph{Thus, our model captures the saturation of the drag coefficient in accordance with  the experimental results of \cite{donelan2004}}.

The theoretical values of the drag coefficient drawn from our model are compared with the original experimental results reported in \cite{donelan2004} and given by the broken line with empty dots in figure \ref{fig:cd}. However, due to an error in processing the experimental data, the corrected results of the same experiment of Donelan et al., were recently published by \cite{CurcicHaus} and are also given in figure \ref{fig:cd} in the solid line with full dots. We also compare our model with the current implementation of $C_D$ in the Geophysical Fluid Dynamics Laboratory hurricane model, recently reported by \cite{Zhouetal2022} and given by the doted line in figure \ref{fig:cd}.

From figure \ref{fig:cd}, one can see that our model predicts a range of values for $C_D$ which is in fairly good agreement with experimental measurements. Moreover, for very strong winds the hypothesis of the K.H. Layer seems valid and even the apparent discrepancy between different results is accounted for by choosing different values of $\rho_d$. 

According to the results reported by \cite{CurcicHaus} for $U_{10} >$ 30 ms$^{-1}$ the effect of droplets account for an eight-fold increase of the air density near the water surface but, according to the values reported by \cite{Zhouetal2022}, for $U_{10} >$ 50 ms$^{-1}$ the effect of droplets should be ignored. 

Discrepancies between different experimental measurements of $C_D$, indicate that the spectral conditions in the vicinity of the measurement point can be different for the same wind speed $U_{10}$ and our model suggests that such environmental conditions are critical for the determination of $\rho_d$ and the exact drag.

Specific values of all the parameters of the model,  for a given $U_{10}$ and a given $\rho_d$, can be obtained by solving numerically the following equation for $U_0$
\begin{align}\label{eq:U0 from U_10}
U_{10} = \frac{u_*(\rho_d,U_0)}{\kappa}\ln\left(\frac{10}{z_0(\rho_d,U_0)}\right) + U_0.
\end{align}
Resulting values of $C_D$, as well as other parameters of the wind and the wave, for $U_{10} = 20, 30, 40, 50 ,60$ ms$^{-1}$ are tabulated in appendix \ref{appA} in tables \ref{table:U10=20} - \ref{table:U10=60}. 

\begin{table}
\centering
\begin{tabular}{cc|ccccccc|}
\cline{3-9}
& & \multicolumn{7}{c|}{$U_{10}$}\\
&  & 20 & 30 & 40 & 50 & 60 & 80 & 100 \\
\cline{1-9}
\multicolumn{1}{|c}{\multirow[c]{5}{*}{$\rho_d$}} & 0.0012 & 1.55 & 1.86 & 1.96 & \textbf{1.98} & 1.96 & 1.89 & 1.81 \\
\multicolumn{1}{|c}{}& 0.0024 & 2.12 & 2.23 & \textbf{2.37} & 2.35 & 2.31 & 2.21 & 2.10 \\
\multicolumn{1}{|c}{}& 0.0048 & 2.61 & \textbf{2.73} & 2.72 & 2.67 & 2.61 & 2.48 & 2.36 \\
\multicolumn{1}{|c}{}& 0.0096 & 3.01 & \textbf{3.05} & 3.00 & 2.93 & 2.86 & 2.72 & 2.57 \\
\multicolumn{1}{|c}{}& 0.0192 & \textbf{3.32} & 3.31 & 3.24 & 3.15 & 3.07 & 2.90 & 2.76 \\
\cline{1-9}
\end{tabular}
\caption{Values of $C_D\times 10^{3}$ as a function of $U_{10}$ in ms$^{-1}$ and $\rho_d$. The largest value in each row is in bold.}\label{table2}
\end{table}

At first glance,  the wind model given by equation \eqref{U} resembles the following logarithmic wind profile which is often used in wave forecasting models:
\begin{align}\label{U:MikyAgnonJanssen}
U_\alpha(z) = \frac{u_*}{\kappa}\ln\left(1 + \frac{z}{z_\alpha}\right),
\end{align}
see equations (2.31) and (2.32) in \cite{komenetal} page  83, with $z_\alpha $ given by Charnok's relation with constant close to $\tilde{\alpha} = 0.0144$
\begin{align}
z_\alpha = \tilde{\alpha}\frac{u_*^2}{g}.
\end{align}
However this model \emph{does not} capture the saturation of the drag coefficient.  Instead, it predicts that $C_D$ grows monotonically as $U_{10}$ increases. 

\section{Concluding remarks}

In the present article we studied the saturation of the drag coefficient at high wind speeds based on the Kelvin Helmholtz instability. This study follows some previous authors and uses the term ``drag saturation'' in its title. However Table \ref{table2}, in which we have highlighted the largest values of $C_D$ in each line, as well as figure \ref{fig:cd}, indicate that the term ``drag maximizing'' would probably be more adequate.

Our model has one basic assumption which is the very existence of the K.H. Layer. This layer is made of an air and droplets mixture. It is characterized by three constants: (i) A thickness $z_0$. (ii) a uniform density $\rho_d$. (iii) A uniform velocity $U_0$. All three parameters relate in a way to the prevailing environmental conditions.

Given the narrowness, of less than two centimeters, of the K.H. Layer and its rather complex surroundings, its physical existence might be difficult to detect, but its phenomenological value has been demonstrated herein. 
Moreover, equation \eqref{U} enables us to find the frictional velocity $u_*$ from any two measured values of a wind profile, say $U_{10}$ and $U_h = U(z = h)$, given by 
\begin{align}
\frac{u_*}{\kappa} = \frac{U_{10} - U_h}{\ln(10/h)}.
\end{align}
Once we know $u_*$ we can calculate $C_D$ from equation \eqref{cd} and obtain $\rho_d$ by interpolating the values in Table 1. Having $U_{10}$ and $\rho_d$ one can get $U_0$, as well as other interesting values, by interpolating in tables \ref{table:U10=20} - \ref{table:U10=60}.

It is quite clear that, under strong winds, a certain part of the gravity waves break, releasing a lot of droplets into the air. These droplets fall  until they reach the K.H. Layer and contribute to the increase in its $\rho_d$. The increase of $\rho_d$ in its turn, as well as the mechanism of the K.H. Layer, are both responsible for the increase in $C_D$. Specifically, the resistance that the gravity wave spectrum applies on the wind is represented by the fact that $\rho_d$ in the K.H. Layer is larger than $\rho_d = 0.0012$, and in the examples herein reaches values up to 0.0192. This phenomenon is demonstrated in tables \ref{table:U10=20} - \ref{table:U10=60} which are given in the appendix \ref{appA}. For $U_{10} = $ 20 ms$^{-1}$ one can see that the presence of a gravity wave spectrum increases $C_D$ from 1.55$\times 10 ^{-3}$ to 3.22$\times 10 ^{-3}$ whereas for $U_{10}$ = 50 ms$^{-1}$ the increase is from 1.98$\times 10 ^{-3}$ to 3.15$\times 10 ^{-3}$.

Unfortunately all laboratory measured droplet mass concentrations, see \cite{Veron2012,Mehta2019}, seem to be given for $z$ values larger than those relevant for the K.H. Layer mentioned herein.




Last, we have tried to include viscous effects into the two fluid model given in section 2. In fact we adapted the ideas of the paper by \cite{DDZ} to our two fluid model, resulting in the following new dispersion relation:
\begin{align}
\omega = \frac{\rho_d kU_0}{1 + \rho_d} \pm \left[ \frac{gk(1 - \rho_d)}{1 + \rho_d} + \frac{\tau k^3}{1 + \rho_d} - \frac{\rho_d k^2U_0^2}{(1 + \rho_d)^2} -\frac{4\tilde{\nu}^2k^4}{1 + \rho_d}\right]^{1/2},
\end{align}
which turns out almost identical to equation \eqref{omega} except for the additional last term under the square root. This term is proportional to $\tilde{\nu}^2$ with $\tilde{\nu} = 10^{-6}$ m$^{2}$s$^{-1}$ being the kinematic viscosity of the water. However the plots of $C_D$ were virtually unaffected by the inclusion of viscosity hence we opted for the inviscid model for our computations.


\appendix\label{appA}
\section{Values of $C_D$ and other parameters for $U_{10} = $ 20,30,40,50 and 60 ms$^{-1}$ and $\rho_d = $ 0.0012,0.0024, 0.0048, 0.0096 and 0.0192}

\begin{table}[H]
\centering
\begin{tabular}{|c|ccccc|}
$\rho_d$ & 0.00120000 & 0.00240000 & 0.00480000 & 0.00960000 & 0.01920000 \\ 
$U_0$ & 7.25391253 & 5.11707499 & 3.58116224 & 2.50004897 & 1.74800569 \\ 
$k_1$ & 204.36711960 & 206.37666684 & 215.39683288 & 228.47513433 & 243.18429757 \\ 
$k_2$ & 647.89525217 & 640.81568634 & 612.50317137 & 574.65736613 & 534.66556670 \\ 
$k_m$ & 475.31904776 & 471.24973521 & 455.35596769 & 434.75608901 & 413.83770156 \\ 
$c_m$ & 0.00869426 & 0.01225158 & 0.01710746 & 0.02377226 & 0.03292946 \\ 
$a_m$ & 2.11019263 & 2.11359792 & 2.12742418 & 2.14687603 & 2.16869999 \\ 
$\chi_m$ & 1.00301475 & 0.99603246 & 0.96873530 & 0.93336742 & 0.89748982 \\ 
$z_0$ & 1.32188797 & 1.33330267 & 1.37984033 & 1.44522077 & 1.51827281 \\ 
$\alpha_m$ & 4.13254830 & 5.77355204 & 7.78998534 & 10.33513322 & 13.62745354 \\ 
$\beta_m$ & 40.53074887 & 39.52838919 & 35.56096481 & 30.32525290 & 24.89099215 \\ 
$\sigma_m$ & 0.74584188 & 1.01952281 & 1.25377140 & 1.44455620 & 1.59534961 \\ 
$u_*$ & 0.78837485 & 0.92173876 & 1.02215923 & 1.09717676 & 1.15302125 \\ 
$C_D$ & 0.00155384 & 0.00212401 & 0.00261202 & 0.00300949 & 0.00332365 \\ 
\end{tabular}
\caption{Resulting parameters of the waves and the wind profile obtained for the wind velocity $U_{10} = 20$ ms$^{-1}$.  All units are in m.k.s., except for $a_m$ and $z_0$ which are in mm and cm respectively.}
\label{table:U10=20}
\end{table}

\begin{table}[H]
\centering
\begin{tabular}{|c|ccccc|}
$\rho_d$ & 0.00120000 & 0.00240000 & 0.00480000 & 0.00960000 & 0.01920000 \\ 
$U_0$ & 8.12240589 & 5.65168609 & 3.90409150 & 2.69179163 & 1.85970528 \\ 
$k_1$ & 143.06863503 & 149.63220356 & 160.15123753 & 173.26035481 & 187.69111875 \\ 
$k_2$ & 925.48926928 & 883.82983249 & 823.79159398 & 757.78973823 & 692.74599212 \\ 
$k_m$ & 643.81816677 & 617.59610751 & 580.16028169 & 539.59262910 & 500.33451398 \\ 
$c_m$ & 0.00973520 & 0.01353157 & 0.01865012 & 0.02559548 & 0.03503369 \\ 
$a_m$ & 1.97808203 & 1.99972788 & 2.02938685 & 2.06027852 & 2.09014391 \\ 
$\chi_m$ & 1.27352515 & 1.23502415 & 1.17736965 & 1.11171110 & 1.04577114 \\ 
$z_0$ & 0.97592544 & 1.01736155 & 1.08300852 & 1.16443127 & 1.25579690 \\ 
$\alpha_m$ & 6.26770172 & 8.35704548 & 10.82005807 & 13.81113397 & 17.52856644 \\ 
$\beta_m$ & 81.92538212 & 75.36770969 & 66.12199017 & 56.22588247 & 46.74729661 \\ 
$\sigma_m$ & 2.00916401 & 2.51871996 & 2.94649151 & 3.29622984 & 3.57977126 \\ 
$u_*$ & 1.29394874 & 1.44876958 & 1.56697466 & 1.65736484 & 1.72717767 \\ 
$C_D$ & 0.00186034 & 0.00233215 & 0.00272823 & 0.00305206 & 0.00331460 \\
\end{tabular}
\caption{Resulting parameters of the waves and the wind profile obtained for the wind velocity $U_{10} = 30$ ms$^{-1}$.  All units are in m.k.s., except for $a_m$ and $z_0$ which are in mm and cm respectively.}
\label{table:U10=30}
\end{table}

\begin{table}[H]
\centering
\begin{tabular}{|c|ccccc|}
$\rho_d$ & 0.00120000 & 0.00240000 & 0.00480000 & 0.00960000 & 0.01920000 \\ 
$U_0$ & 9.00682050 & 6.19590298 & 4.23468236 & 2.89154818 & 1.98037392 \\ 
$k_1$ & 109.97839236 & 117.61111590 & 128.15306989 & 140.60988847 & 153.97783839 \\ 
$k_2$ & 1203.95000912 & 1124.46348623 & 1029.48172332 & 933.75309762 & 844.42197415 \\ 
$k_m$ & 822.27670843 & 770.86300389 & 709.79947992 & 648.78519052 & 592.44400475 \\ 
$c_m$ & 0.01079523 & 0.01483456 & 0.02022937 & 0.02749491 & 0.03730689 \\ 
$a_m$ & 1.81578776 & 1.86432821 & 1.92054710 & 1.97389899 & 2.01980612 \\ 
$\chi_m$ & 1.49307998 & 1.43714165 & 1.36320333 & 1.28063643 & 1.19662203 \\ 
$z_0$ & 0.76412055 & 0.81508456 & 0.88520568 & 0.96845387 & 1.06055345 \\ 
$\alpha_m$ & 8.87666645 & 11.43541672 & 14.35879938 & 17.83829132 & 22.10224156 \\ 
$\beta_m$ & 128.54088418 & 114.65175146 & 98.58995495 & 82.98439006 & 68.93807249 \\ 
$\sigma_m$ & 3.76202218 & 4.55698326 & 5.22156537 & 5.76765809 & 6.21605468 \\ 
$u_*$ & 1.77059834 & 1.94871395 & 2.08597806 & 2.19234617 & 2.27597135 \\ 
$C_D$ & 0.00195939 & 0.00237343 & 0.00271957 & 0.00300399 & 0.00323753 \\ 
\end{tabular}
\caption{Resulting parameters of the waves and the wind profile obtained for the wind velocity $U_{10} = 40$ ms$^{-1}$.  All units are in m.k.s., except for $a_m$ and $z_0$ which are in mm and cm respectively.}
\label{table:U10=40}
\end{table}

\begin{table}[H]
\centering
\begin{tabular}{|c|ccccc|}
$\rho_d$ & 0.00120000 & 0.00240000 & 0.00480000 & 0.00960000 & 0.01920000 \\ 
$U_0$ & 9.91064965 & 6.74962643 & 4.56925792 & 3.09284650 & 2.10192329 \\ 
$k_1$ & 88.11061541 & 95.96978041 & 106.26580733 & 118.17980799 & 130.82055086 \\ 
$k_2$ & 1502.75294149 & 1378.03175996 & 1241.52111165 & 1110.97590321 & 993.89789611 \\ 
$k_m$ & 1017.18521740 & 935.54754971 & 846.57772108 & 762.00302075 & 686.69752763 \\ 
$c_m$ & 0.01187853 & 0.01616032 & 0.02182767 & 0.02940900 & 0.03959667 \\ 
$a_m$ & 1.63344375 & 1.70856885 & 1.79269450 & 1.87260901 & 1.94114848 \\ 
$\chi_m$ & 1.66151483 & 1.59844740 & 1.51765522 & 1.42693372 & 1.33298186 \\ 
$z_0$ & 0.61770317 & 0.67160513 & 0.74218647 & 0.82456173 & 0.91498586 \\ 
$\alpha_m$ & 12.08266040 & 15.11874653 & 18.47881506 & 22.40974683 & 27.19093612 \\ 
$\beta_m$ & 184.16439007 & 160.17975236 & 135.01569572 & 112.02080550 & 92.27182360 \\ 
$\sigma_m$ & 5.93713047 & 7.06949468 & 8.01809042 & 8.80298293 & 9.45389506 \\ 
$u_*$ & 2.22432208 & 2.42718882 & 2.58490658 & 2.70847173 & 2.80682131 \\ 
$C_D$ & 0.00197904 & 0.00235650 & 0.00267270 & 0.00293433 & 0.00315130 \\
\end{tabular}
\caption{Resulting parameters of the waves and the wind profile obtained for the wind velocity $U_{10} = 50$ ms$^{-1}$.  All units are in m.k.s., except for $a_m$ and $z_0$ which are in mm and cm respectively.}
\label{table:U10=50}
\end{table}

\begin{table}[H]
\centering
\begin{tabular}{|c|ccccc|}
$\rho_d$ & 0.00120000 & 0.00240000 & 0.00480000 & 0.00960000 & 0.01920000 \\ 
$U_0$ & 10.83792400 & 7.31610636 & 4.90998876 & 3.29668184 & 2.22429052 \\ 
$k_1$ & 72.34905063 & 80.06691079 & 89.97458947 & 101.37545661 & 113.45441710 \\ 
$k_2$ & 1830.13440172 & 1651.73608059 & 1466.31670134 & 1295.13516690 & 1146.03092231 \\ 
$k_m$ & 1232.51238793 & 1114.99884247 & 993.25165739 & 881.35035656 & 784.40377162 \\ 
$c_m$ & 0.01298992 & 0.01751662 & 0.02345536 & 0.03134721 & 0.04190186 \\ 
$a_m$ & 1.45190316 & 1.54766639 & 1.65517824 & 1.75966319 & 1.85161407 \\ 
$\chi_m$ & 1.78948863 & 1.72564624 & 1.64400853 & 1.55087978 & 1.45241306 \\ 
$z_0$ & 0.50978679 & 0.56351496 & 0.63258745 & 0.71290438 & 0.80101416 \\ 
$\alpha_m$ & 16.01023842 & 19.53100589 & 23.29707551 & 27.62787687 & 32.86797880 \\ 
$\beta_m$ & 251.31820418 & 213.83032503 & 176.79770258 & 144.39197833 & 117.55236566 \\ 
$\sigma_m$ & 8.48197617 & 10.00342250 & 11.28411651 & 12.35235254 & 13.24662721 \\ 
$u_*$ & 2.65863000 & 2.88724530 & 3.06650133 & 3.20836829 & 3.32247739 \\ 
$C_D$ & 0.00196342 & 0.00231561 & 0.00261206 & 0.00285934 & 0.00306635 \\ 
\end{tabular}
\caption{Resulting parameters of the waves and the wind profile obtained for the wind velocity $U_{10} = 60$ ms$^{-1}$.  All units are in m.k.s., except for $a_m$ and $z_0$ which are in mm and cm respectively.}
\label{table:U10=60}
\end{table}


\bibliographystyle{plainnat}
\bibliography{References}

\begin{thebibliography}{13}
\providecommand{\natexlab}[1]{#1}
\providecommand{\url}[1]{\texttt{#1}}
\expandafter\ifx\csname urlstyle\endcsname\relax
  \providecommand{\doi}[1]{doi: #1}\else
  \providecommand{\doi}{doi: \begingroup \urlstyle{rm}\Url}\fi

\bibitem[Curcic and Haus(2020)]{CurcicHaus}
M.~Curcic and B.~K. Haus.
\newblock Revised estimates of ocean surface drag in strong winds.
\newblock \emph{Geophysical Research Letters}, 47\penalty0 (10):\penalty0
  e2020GL087647, 2020.

\bibitem[Dias et~al.(2008)Dias, Dyachenko, and Zakharov]{DDZ}
F.~Dias, A.~I. Dyachenko, and V.~E. Zakharov.
\newblock Theory of weakly damped free-surface flows: A new formulation based
  on potential flow solutions.
\newblock \emph{Physics Letters A}, 372\penalty0 (8):\penalty0 1297--1302,
  2008.
\newblock ISSN 0375-9601.
\newblock \doi{https://doi.org/10.1016/j.physleta.2007.09.027}.
\newblock URL
  \url{https://www.sciencedirect.com/science/article/pii/S037596010701345X}.

\bibitem[Donelan et~al.(2004)Donelan, Haus, Reul, Plant, Stiassnie, Graber,
  Brown, and Saltzman]{donelan2004}
M.~A. Donelan, B.~K. Haus, N.~Reul, W.J. Plant, M.~Stiassnie, H.C. Graber, O.B.
  Brown, and E.S. Saltzman.
\newblock On the limiting aerodynamic roughness of the ocean in very strong
  winds.
\newblock \emph{Geophysical Research Letters}, 31\penalty0 (18), 2004.

\bibitem[Hogan(1980)]{Hogan1980b}
S.J. Hogan.
\newblock Some effects of surface tension on steep water waves. part 2.
\newblock \emph{Journal of Fluid Mechanics}, 96\penalty0 (3):\penalty0
  417--445, 1980.

\bibitem[Iida et~al.(1992)Iida, Toba, and Chaen]{Iida1992}
N.~Iida, Y.~Toba, and M.~Chaen.
\newblock A new expression for the production rate of sea water droplets on the
  sea surface.
\newblock \emph{Journal of Oceanography}, 48\penalty0 (4):\penalty0 439--460,
  1992.

\bibitem[Komen et~al.(1996)Komen, Cavaleri, Donelan, Hasselmann, Hasselmann,
  and Janssen]{komenetal}
G.~J. Komen, L.~Cavaleri, M.~Donelan, K.~Hasselmann, S.~Hasselmann, and
  P.A.E.M. Janssen.
\newblock \emph{Dynamics and modelling of ocean waves}.
\newblock Cambridge University Press, 1996.

\bibitem[Mehta et~al.(2019)Mehta, Ortiz-Suslow, Smith, and Haus]{Mehta2019}
S.~Mehta, D.~G. Ortiz-Suslow, A.~W. Smith, and B.~K. Haus.
\newblock A laboratory investigation of spume generation in high winds for
  fresh and seawater.
\newblock \emph{Journal of Geophysical Research: Atmospheres}, 124\penalty0
  (21):\penalty0 11297--11312, 2019.

\bibitem[Phillips(1966)]{phillipsbook}
O.~M. Phillips.
\newblock \emph{The Dynamics of the Upper Ocean}.
\newblock Cambridge University Press, 1966.
\newblock ISBN 9780521214216.

\bibitem[Soloviev and Lukas(2010)]{SolovievLukas2010}
A.~Soloviev and R.~Lukas.
\newblock Effects of bubbles and sea spray on air--sea exchange in hurricane
  conditions.
\newblock \emph{Boundary-layer meteorology}, 136\penalty0 (3):\penalty0
  365--376, 2010.

\bibitem[Stiassnie et~al.(2007)Stiassnie, Agnon, and Janssen]{MikyAgnonJanssen}
M~Stiassnie, Y~Agnon, and PAEM Janssen.
\newblock Temporal and spatial growth of wind waves.
\newblock \emph{Journal of physical oceanography}, 37\penalty0 (1):\penalty0
  106--114, 2007.

\bibitem[Troitskaya and Rybushkina(2008)]{TroitskayaRybushkina2008}
Yu.~I Troitskaya and G.~V. Rybushkina.
\newblock Quasi-linear model of interaction of surface waves with strong and
  hurricane winds.
\newblock \emph{Izvestiya, Atmospheric and Oceanic Physics}, 44:\penalty0 621
  -- 645, 2008.
\newblock \doi{10.1134/S0001433808050083}.
\newblock URL \url{https://doi.org/10.1134/S0001433808050083}.

\bibitem[Veron et~al.(2012)Veron, Hopkins, Harrison, and Mueller]{Veron2012}
F.~Veron, C.~Hopkins, E.L. Harrison, and J.A. Mueller.
\newblock Sea spray spume droplet production in high wind speeds.
\newblock \emph{Geophysical Research Letters}, 39\penalty0 (16), 2012.

\bibitem[Zhou et~al.(2022)Zhou, Hara, Ginis, D’Asaro, Hsu, and
  Reichl]{Zhouetal2022}
X.~Zhou, T.~Hara, I.~Ginis, E.~D’Asaro, J.~Hsu, and B.G. Reichl.
\newblock Drag coefficient and its sea state dependence under tropical
  cyclones.
\newblock \emph{Journal of Physical Oceanography}, 52\penalty0 (7):\penalty0
  1447 -- 1470, 2022.
\newblock \doi{10.1175/JPO-D-21-0246.1}.
\newblock URL
  \url{https://journals.ametsoc.org/view/journals/phoc/52/7/JPO-D-21-0246.1.xml}.

\end{thebibliography}

\end{document}